%% file: sigconf.tex
\documentclass[sigconf]{acmart}

\usepackage{booktabs} 
\usepackage{todonotes} 
\usepackage{lipsum} 

\usepackage[utf8]{inputenc}
\usepackage[T1]{fontenc}

\usepackage{subfigure}
\usepackage{amsmath}
\usepackage{balance}

\begin{document}

\copyrightyear{2019}
\acmYear{2019}
\setcopyright{acmlicensed}
\acmConference[SAC '19]{The 34th ACM/SIGAPP Symposium on Applied Computing}{April 8--12, 2019}{Limassol, Cyprus}
\acmBooktitle{The 34th ACM/SIGAPP Symposium on Applied Computing (SAC '19), April 8--12, 2019, Limassol, Cyprus}
\acmPrice{15.00}
\acmDOI{10.1145/3297280.3299802}
\acmISBN{978-1-4503-5933-7/19/04}

\title{From spatio-temporal data to chronological networks: An application to wildfire analysis}

\author{Didier A. Vega-Oliveros}
\orcid{0000-0001-9569-3775}
 \affiliation{%
  \institution{Department of Computing and Mathematics, University of S\~ao Paulo}
  \city{Ribeir\~ao Preto}
  \state{SP}
  \country{Brazil}
 }
\email{davo@icmc.usp.br}

\author{Mosh\'e Cotacallapa}
 \affiliation{%
  \institution{National Institute for Space Research}
  \city{S\~ao Jos\'e dos Campos}   
  \state{SP}
  \country{Brazil}
 }
\email{frank.moshe@inpe.br}

\author{Leonardo N. Ferreira}
\affiliation{ 
  \institution{National Institute for Space Research}
  \city{S\~ao Jos\'e dos Campos}
  \state{SP}
  \country{Brazil}
  }
 \email{leonardo.ferreira@inpe.br}
 
\author{Marcos Quiles}
 \affiliation{
 \institution{Institute of Science and Technology, Federal University of S\~ao Paulo}
  \city{S\~ao Jos\'e dos Campos}
  \state{SP}
  \country{Brazil}
 }
 \email{quiles@unifesp.br}
 
\author{Liang Zhao}
 \affiliation{
 \institution{Department of Computing and Mathematics, University of S\~ao Paulo}
  \city{Ribeir\~ao Preto}
  \state{SP}
  \country{Brazil}
 }
 \email{zhao@usp.br}
 
\author{Elbert E. N. Macau}
 \affiliation{
 \institution{Institute of Science and Technology, Federal University of S\~ao Paulo}
  \city{S\~ao Jos\'e dos Campos}
  \state{SP}
  \country{Brazil}
 }
 \email{elbert.macau@inpe.br} 
 
\author{Manoel F. Cardoso}
 \affiliation{
 \institution{Center for Earth System Science, National Institute for Space Research}
  \city{Cachoeira Paulista}
  \state{SP}
  \country{Brazil}
 }
 \email{manoel.cardoso@inpe.br}

\renewcommand{\shortauthors}{D. A. Vega-Oliveros et al.}

\begin{abstract}

Network theory has established itself as an appropriate tool for complex systems analysis and pattern recognition. In the context of spatiotemporal data analysis, correlation networks are used in the vast majority of works. However, the Pearson correlation coefficient captures only linear relationships and does not correctly capture recurrent events. This missed information is essential for temporal pattern recognition. In this work, we propose a chronological network construction process that is capable of capturing various events. Similar to the previous methods, we divide the area of study into grid cells and represent them by nodes. In our approach, links are established if two consecutive events occur in two different nodes. Our method is computationally efficient, adaptable to different time windows and can be applied to any spatiotemporal data set. As a proof-of-concept, we evaluated the proposed approach by constructing chronological networks from the MODIS dataset for fire events in the Amazon basin. We explore two data analytic approaches: one static and another temporal. The results show some activity patterns on the fire events and a displacement phenomenon over the year. The validity of the analyses in this application indicates that our data modeling approach is very promising for spatio-temporal data mining.

\end{abstract}



\maketitle

\input{body-conf}

\bibliographystyle{ACM-Reference-Format}
\bibliography{bibliography} 
\balance

\end{document}

%% file: body-conf.tex
\section{Introduction}

Every second, a vast amount of spatio-temporal data are produced in the whole world. Examples include phone calls, live concerts, crimes, and car accidents. The data collected from these events are usually used to understand the behavior and to get much other insightful information from the complex systems they are part of. In this sense, Network Sciences have provided several resources for this purpose. The network representation allows the study of the interactions and dynamics between the small components from the represented complex systems in a unified way.

A common approach to represent geographical data into networks is based on the construction process of linking nodes according to their correlation coefficients, which are calculated from the underlying time series for each point of the spatial grid. This approach has been successfully used in a wide range of scientific areas. For instance, in Earth Sciences, networks have been used to analyze global climate \cite{Zhou2015}, to predict El-Ni\~no, and explore its impact around the world \cite{Tsonis2008, Meng2018,  Fan2017}. In Bioinformatics, networks were applied to study gene expression \cite{Farkas2003}. In Finance, it has been used to understand and find the dynamics of financial markets \cite{Bialonski2011}.

Network science has helped to identify valuable information in many domains. However, several questions have raised around the limitations and applications of correlation networks. For example, what is the minimum time series length to consider the correlation? Is this long-length enough to find statistically significant correlations? What is the minimum correlation threshold to connect two nodes? Which temporal patterns can and cannot be captured by this construction process?
Notwithstanding having answers to all of these questions, the correlation-based networks are not appropriate to many real-world spatio-temporal systems. For instance, if we only have short-length time series, how can we get a precise correlation coefficient between those time series? How can we measure how much the system changed between short time intervals?

To tackle the above questions, we propose an alternative approach to overcome those constraints: The chronological network construction. Similar to the previous works, our method represents grid cells of a geographical region by nodes. However, we connect them in a different way. A link is created between nodes if two consecutive events occur between them. If those events are in the same grid point, the nodes form a self-loop. Although being easy to build and use, very few studies have explored the potential of this network construction process to model and analyze spatio-temporal data sets. 
In this paper, we apply our new method to study the Amazon Basin fire activity. The Amazon, a vast region in South-America, is covered by rain-forest that contains a colossal biodiversity~\cite{Vieira2008}. Unfortunately, over the recent decades, the region has been directly affected by deforestation, which is influenced by several factors like urban growth, farming areas, cattle industry, roadworks, fires, among others \cite{Vieira2008, Zemp2017}. The Amazon Basin is the largest drainage basin in the world, discharging about 209,000 cubic meters per second to the ocean~\cite{Vieira2008, Zemp2017}. With that quantity of water flowing through the Amazon region, it seems that fire propagation in large areas, as occurs in other parts of the world, is an unlikely event. However, looking at fire data sets collected from satellite, wildfires in the Amazon are very dense and active in different regions. Although those fires events are not directly related, as a complex system, they share conditions that increase the probability of happening in some areas, even when they are far between each other.


Wildfire has a considerable impact on human life. This environmental process is responsible for vegetation composition changes \cite{Brando2014}, emission of gases and particles into the atmosphere \cite{Gatti2014}, and damage to properties \cite{Dey2018}. Furthermore, climate change has been changing the frequency and severity of fire \cite{AdamRogers}. The many factors that influence fire activity make this complex system hard to forecast. In recent years, complex networks emerged as a powerful tool to study systems like this. Since fires dynamics is still poorly studied using network science, we believe that this application domain suits very well for our method. In summary, our goal in this paper is to use our network construction method to model and study wildfires focusing on the advantages and limitations of our method.

\section{Related work}
\label{sec:related-work}

Concerning spatio-temporal events, \citeauthor{Abe2006}~\cite{Abe2006} employed a similar approach called sequential networks, for describing and finding patterns on the earthquake network of United States of America and Japan. They also analyzed the topological properties on growing networks. \citeauthor{Zemp2017}~\cite{Zemp2017} model the moisture recycling process in South America and developed a similar framework to event-based networks with weighted nodes and directed links. More recently, \citeauthor{Ferreira2018}~\cite{Ferreira2018} reported a variation the work in ~\cite{Abe2006} introducing window times and new rules to link nodes. As results, the authors obtained a better understanding of long-range seismic activities. The before works are interesting due to the simplicity to build the data model network and to find connections between different regions, even from the world~\cite{Ferreira2018}. Therefore, the following sections will introduce the proposed methods to explore the advantages of network characterization based on the chronological approach, with potential applications in several fields.

\section{Background}
	
Let consider a geographical data with a set of spatial points. They can be represented by the network $G=(V,E)$, where $V$ is the set of $n$ spatial located points called as nodes or vertices $V = \{v_1, v_2, \ldots, v_n\}$, and the set of $m$ edges or links $E = \{e_1, e_2, \ldots, e_m\}$ denoting some similarity or distance relationship. The adjacency matrix  $A_{n\times n}$ is the mathematical entity of the network, where $A_{ij} = 1$ means there exists a connection between nodes $v_i$ and $v_j$, and $A_{ij} = 0$, otherwise. The links can be directed, indicating the initial and target node of the relationship, or undirected where both have the same relationship. Also, the elements of $E$ can be weighted, meaning the similarity strength between the nodes. 

The degree of connectivity of node $v_i$, called as $k_i$, is the number of links  incident on $v_i$. When the links are weighted, the alternative is to calculate the strength $s_i$ of the nodes, which is the sum of the weights of incident links of each node. In the case of directed networks, $k_i$ is the sum of the degrees of input (links that reach the node) and output (links that leave the node). 
The degree distribution of a network $P(k)$ is the probability of randomly select a node with degree $k$. The level of disorder or heterogeneity of nodes connections is obtained with the entropy of the degree distribution, calculated by the normalized  Shannon entropy~\cite{wang06}, i.e., \begin{equation}
\label{eq:shannon}
  \tilde{H} = - \frac{\sum_{k = 0}^{\infty} P(k) \log(P(k))}{\log(N)}\: ,
\end{equation} with $ 0 \leq  \tilde{H} \leq 1 $. The more heterogeneous the distribution the greater the entropy, resulting in $1$ with a uniform $P(k)$ and $0$ when all vertices have the same degree. The entropy of a network is related to the robustness and level of resilience~\cite{wang06}.

The network decomposition into shells or K-Core~\cite{kcore:seidman83} (K $ \in \mathbb{N}_{>0}$ or core of order K) is the maximum subset of nodes that have at least degree $k_i \geq$ K and the K-Core is the highest-order core they belong to. Formally, a node $v_i$ is in a core of order K $\iff$ $v_i$ belongs to the K-Core but not the (K+1)-Core decomposition~\cite{kcore:seidman83,Kitsak2010}. Vertices with the highest coreness are the most central. On the other hand, communities are sets of densely interconnected vertices and sparsely connected with the rest of the network~\cite{Newman2010}. Nodes that belong to the same community, in general, share common properties and perform similar roles. Therefore, the division of a network into communities helps to understand their topological structure (structural and functional properties) and its dynamical processes, obtaining relevant information and features to the network domain. Several methods have been reported to detect communities on networks~\cite{Fortunato16}. Two methods adopted here are the agglomerative and optimization fastgreedy algorithm~\cite{newman2004}, and the fast modularity optimization method~\cite{Lambiotte2014}.

\section{Exploring spatiotemporal patterns in complex networks}
In a geographical complex system we have the set of spatial-located points producing some signal or information over time. This information can be represented as chronological events and employed for discovering global, local and intermediate patterns of the studied system. We presented our approach of grouping this points in grid cells and constructing an event-based characteristic network. The construction of this complex network, associated with the geographical system, can be conducted for the whole period of time or in fixed/dynamic intervals following a multigraph or temporal networks technique. Next, we present our event-based modeling network and two data analytic approaches with results and interpretations related to the fire-event Geographical system.

\subsection{Data}
\label{subsec_data}

In a global scale, the fire event activity is collected mainly through satellite instruments like the Moderate Resolution Imaging Spectroradiometer (MODIS) and Visible Infrared Imaging Radiometer Suite (VIIRS). The MODIS runs in both, Aqua and Terra satellites, operated by the National Aeronautics and Space Administration (NASA), while the VIIRS works at Suomi-NPP satellite, also by the NASA and the National Oceanic and Atmospheric Administration (NOAA). In this research, we employ the MODIS data due to the more than 15 years of available fire event around the world and the data confidence for the continuous refinements and calibrations performed in the MODIS system. 

Here, we perform the research using data from the last version (C6) of MODIS. The time interval is between 01 January 2003 and 31 January 2018 and the region under study is a portion of the Amazon basin, that is located between longitude $70^{\circ}$W, $50^{\circ}$W and latitude $15^{\circ}$S, $5^{\circ}$N. From the MODIS data, we consider the UTC date and hour, both satellites Terra and Aqua, the geographic coordinates, and the detection confidence of the fire event. The total number of occurrences is 1,684,600, after a filtering process considering detection confidence above 70\%. It is relevant to notice that the satellites sequentially scan the earth surface, i.e., the capture is by resolution points. Therefore, there are not parallel events at the same time in the data set.

\subsection{Event-based Characteristic Network}
\label{subsec_method}

Our network characterization process for spatio-temporal events is based on these three steps:

\begin{enumerate}
    \item \emph{Grid-division:} A geographical region under consideration is divided in a grid. Each grid cell is represented as a node in the network.
    \item \emph{Time length:} The network data modeling can be defined for specific periods of time, e.g., the whole time period, fixed or dynamic intervals.
    \item \emph{Links construction:} From the data set, two successive events create a link between the grid cells where they are located.
\end{enumerate} 

\begin{figure}[t!b]
\includegraphics[width=0.46\textwidth]{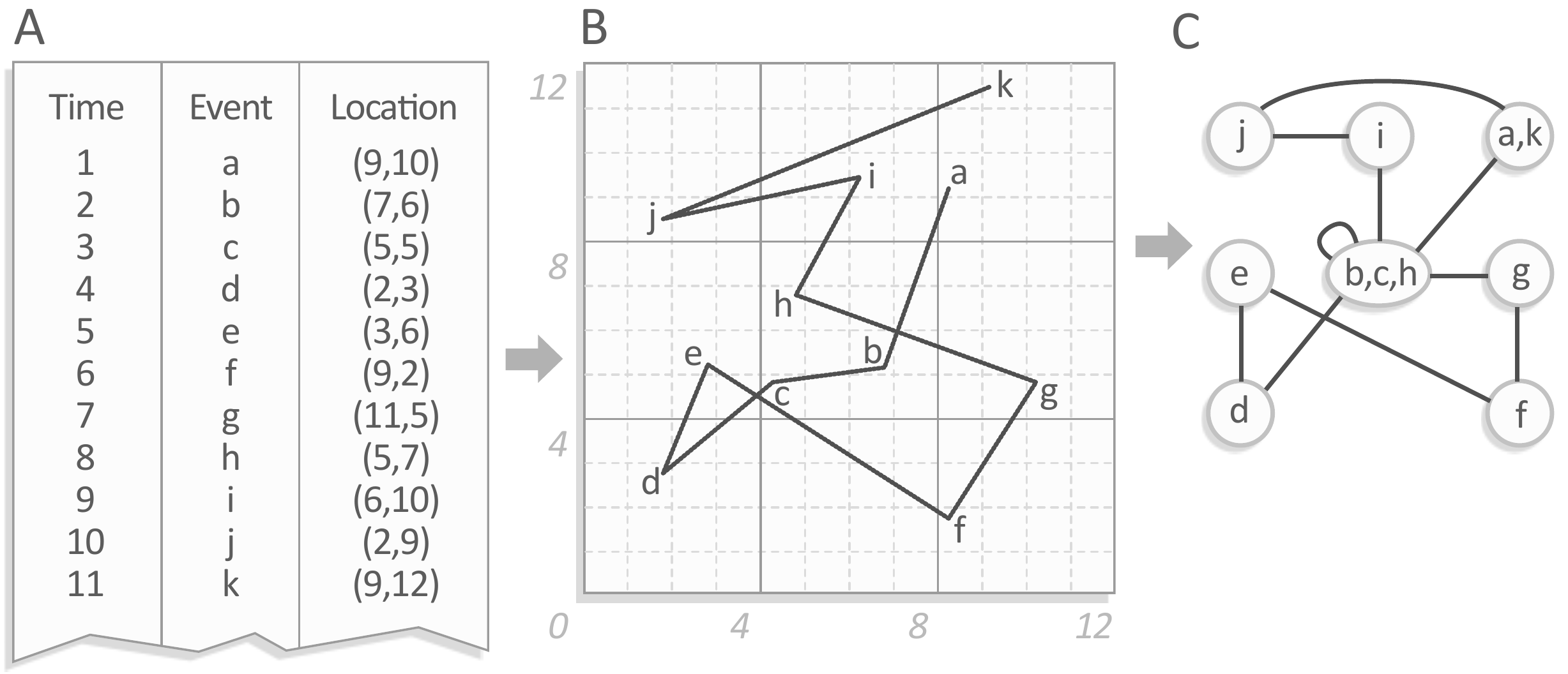}
\caption{\small Illustrative example of a set of spatio-temporal events transformed into a network by following the chronological characterization.}
\label{fig:Spatio-Temporal-Network}
\end{figure}

In Figure~\ref{fig:Spatio-Temporal-Network}, we show a representation of the network construction method and the tackled problem. Every event has a different timestamp, even that wildfires in different geographic areas can occur at the same time (Figure~\ref{fig:Spatio-Temporal-Network}.A). The before is because fire events in the MODIS data are sequential. However, in the case of parallel occurrences, the construction process can proceed branching the connections. The spatio-temporal data is represented in a $4\times4$ grid-division, and the events are linked as they occur (Figure~\ref{fig:Spatio-Temporal-Network}.B). Then, in Figure~\ref{fig:Spatio-Temporal-Network}.C, we have the chronological network representation of the data. Our approach has low code complexity and linear computational cost $(\mathcal{O}(N))$, with $N$ being the number of processed geographical events. Furthermore, the faster network construction allows a real-time data streams ingestion. To make reproducibility easier, we share the implementation of our method online\footnote{Code available at \url{https://github.com/fire-networks}}.

Some points concerning this method are raised here. First, what is the optimal grid-division? It depends on the quantity of data and the spatial distribution, in order to avoid dense networks -- where almost every node is connected to each other (a feasible solution is to increase the grid-division or to create time slices), and also to avoid sparse networks (a solution could be to decrease the grid-division or to increase the time length). Second, what happens if we have self-loops or multiple links between two nodes? For instance, Figure~\ref{fig:Spatio-Temporal-Network}.C we have a self-loop in node $(b,e,h)$. After the linking process, the network can be simplified by removing self-loops, the directions and multiple links. Each type of network can reveal different properties of the characterized system. However, following different networks analysis can lead to similar results (see discussion in Section~\ref{sec:finalRemarks}). 

\begin{figure}[ht]
\includegraphics[width=0.4\textwidth]{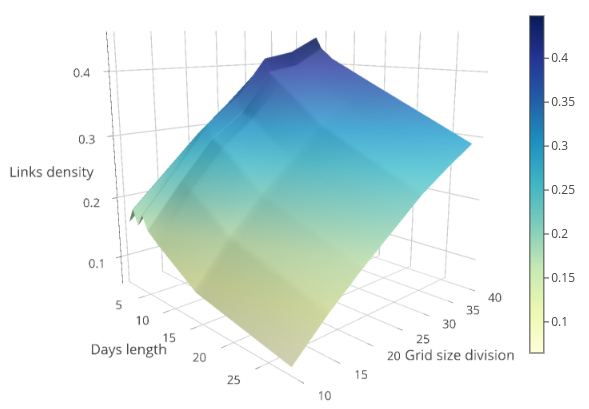}
\caption{\small Sensibility analysis by measuring the links density for grid size division and days length values. The darker the color, the higher the links density.}
\label{fig:sensibility}
\end{figure}

According to the given considerations, we analyze two different network approaches to show the contributions of our chronological method: {\bf (\textit{I})} Building a single network with multiple directed links and no self-loops for the whole period (Section~\ref{subsec_single_net}). Thus, we analyze the long-term properties of the system. {\bf (\textit{II})} Building unweighted and undirected temporal networks from the whole period. Thus, we explore the geographical system evolution over the 15 years (Section~\ref{subsec_temporal}).  The best parameter combination (grid size and time length) was found by performing a sensibility analysis, seeking to maintain a consistent links density after the simplification process. As illustrated in Figure~\ref{fig:sensibility}, an optimal grid-division for the Amazon basin area is $30\times30$, and for the temporal division, periods of seven days. The reason is that at this points we achieve a link density plateau with almost $40\%$ of the original edges after the simplification.

\begin{figure}[b!]
\includegraphics[width=0.4\textwidth]{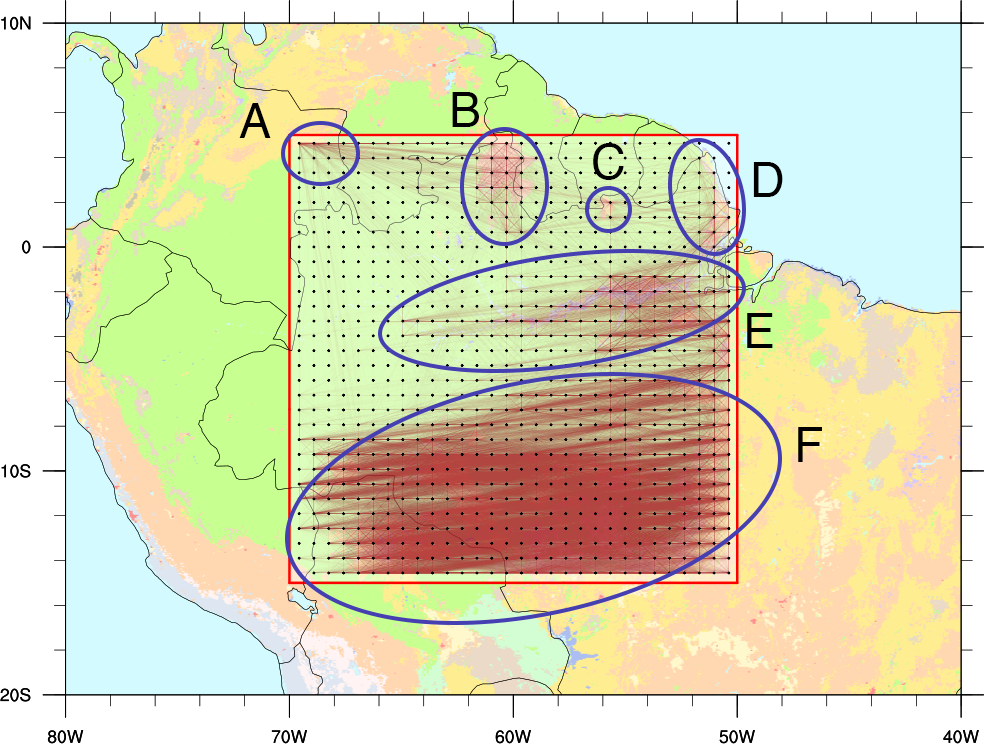}
\caption{\small Network representing the whole data set. The small dots are the nodes (grid cells) and red lines represent the links (consecutive events). The background colors represent different land uses. The line transparency is inversely proportional to the link weight. This network is characterized by six regions with fire activity: (A) east Colombia (B) Roraíma state in Brazil, (C) Park of Tumucumaque in north Pará state in Brazil, (D) Amapá state in Brazil, (E) Amazon river banks, and (F) southeast Amazon forest region.}
\label{fig_leo1}
\end{figure}

\subsection{Single Network Approach}
\label{subsec_single_net}

In this section, we use our method (Sec.~\ref{subsec_method}) to construct a single network and study the historical data (Sec.~\ref{subsec_data}). The goal is to use this network to characterize the whole data set and search for patterns. We start by constructing a directed and weighted network. The link weights represent the frequency of fire occurrences between two nodes on the data set, and the direction is the chronology flow. Then, we simplify it into an undirected and weighted network without self-loops, by adding the weights (in- and out-links) between two nodes, i.e., $\mathbf{A} = \mathbf{A} + \mathbf{A}^\intercal$. Figure~\ref{fig_leo1} illustrates the resulting network that can be divided into six regions of fire activity. For the sake of simplicity, we will refer to this network as ``historical network''.

\begin{figure}[bt!]
\includegraphics[width=0.4\textwidth]{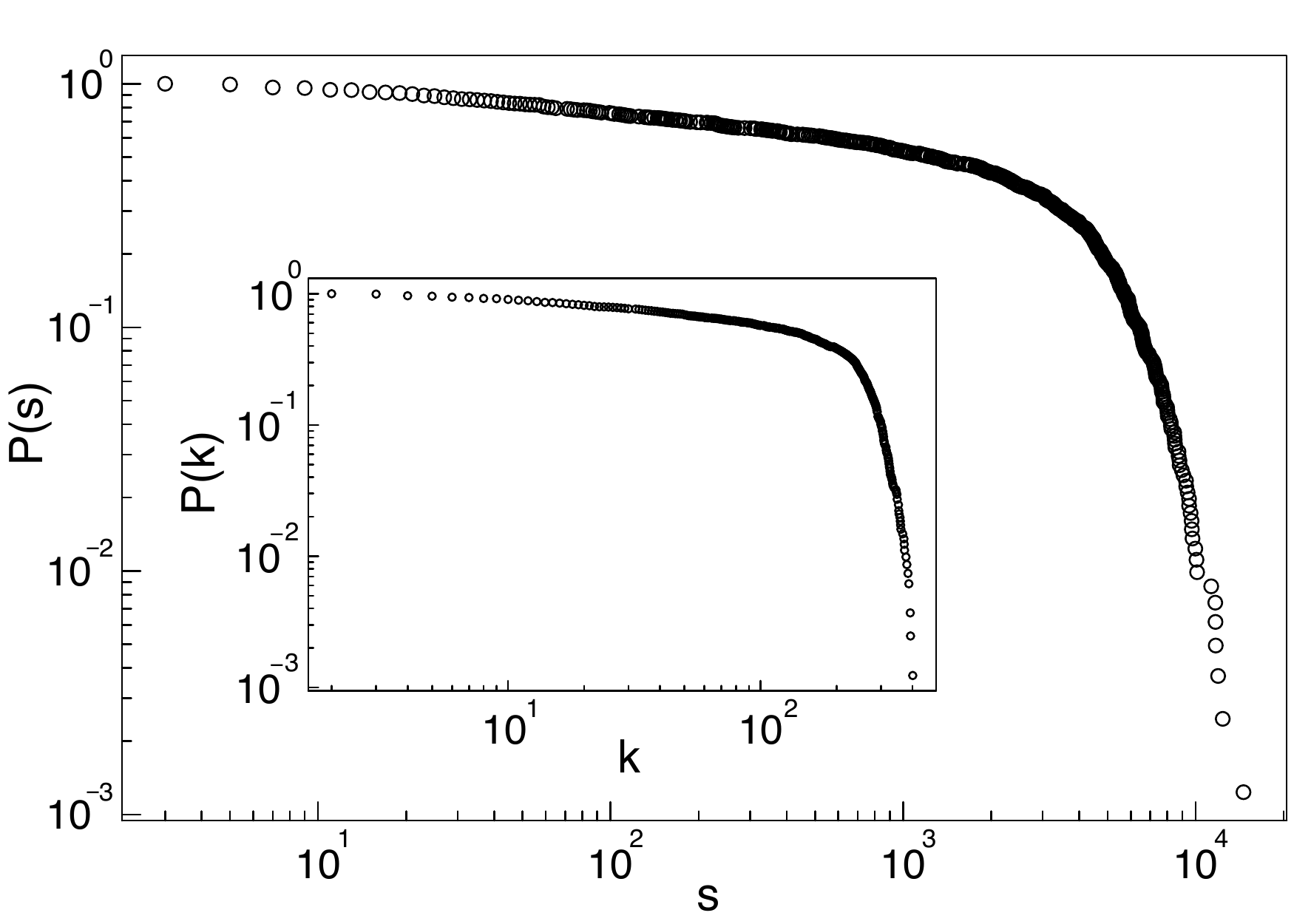}
\caption{\small Strength and degree (inset) cumulative distributions. Both distributions have decay sharper than a power-law. }
\label{fig_leo3}
\end{figure}

\begin{figure}[t!b]
\includegraphics[width=0.4\textwidth]{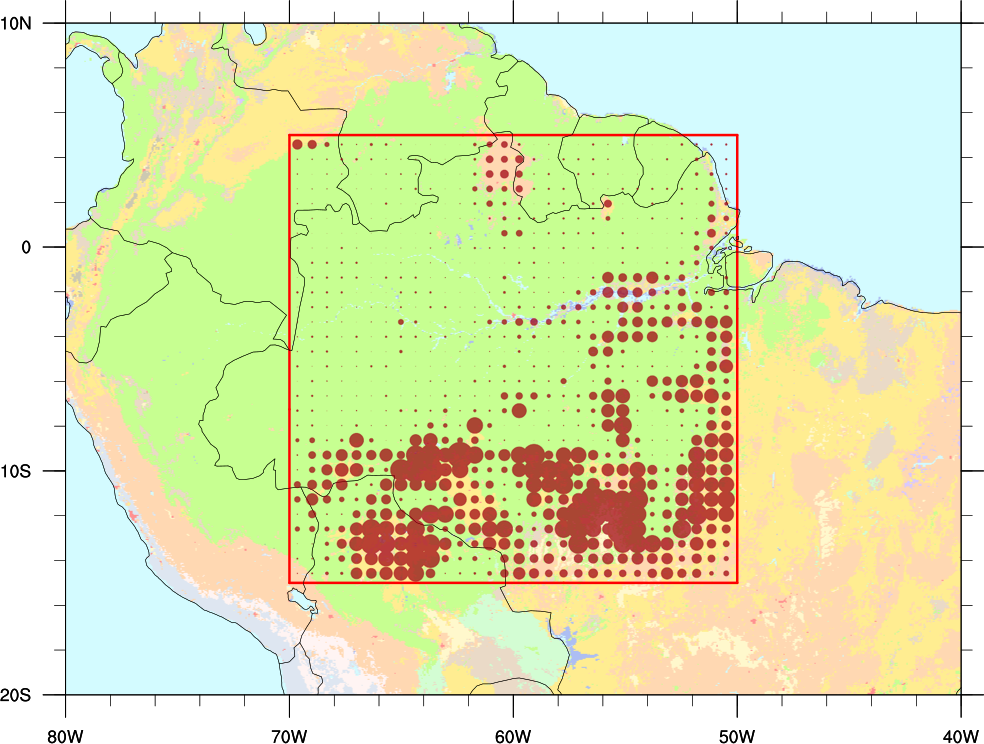}
\caption{\small Nodes strength for the historical network. The sizes of the nodes (red dots) are proportional to the strength.}
\label{fig_leo4}
\end{figure}

\begin{figure}[h]
\includegraphics[width=0.4\textwidth]{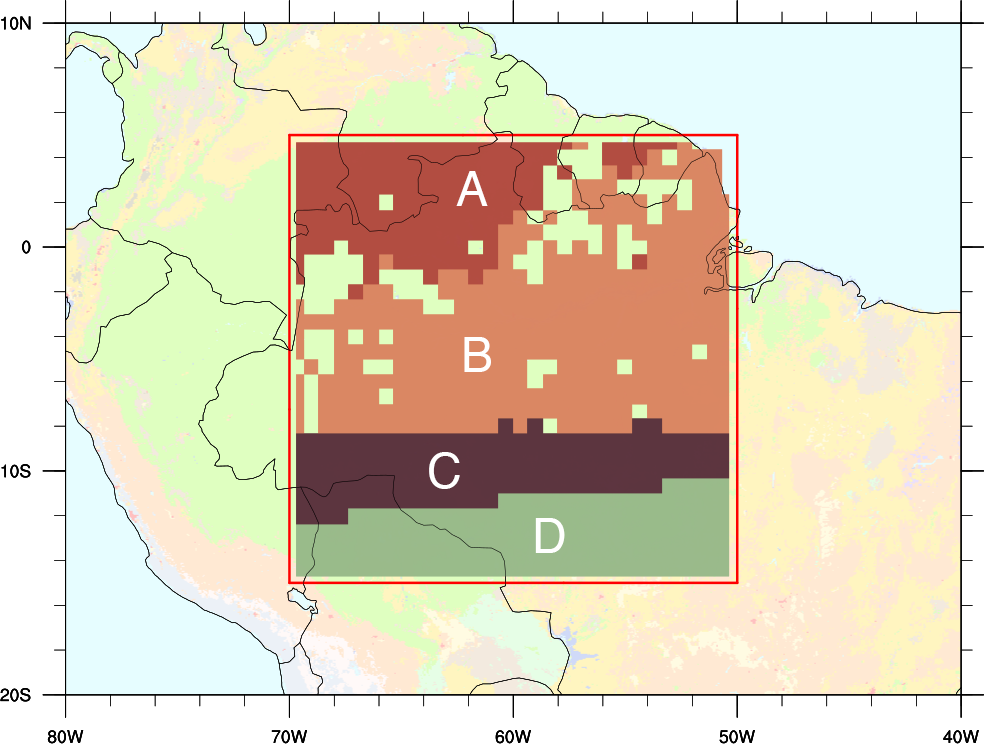}
\caption{\small Community structure found with the fast greedy algorithm (weighted) \cite{newman2004}. The colors represent the four communities achieved by the highest modularity. In our network construction method, a community represents a region of frequent fire activity that occur in a period of time.}
\label{fig_leo2}
\end{figure}

\begin{figure*}[]
\includegraphics[width=0.75\textwidth,height=0.37\textwidth]{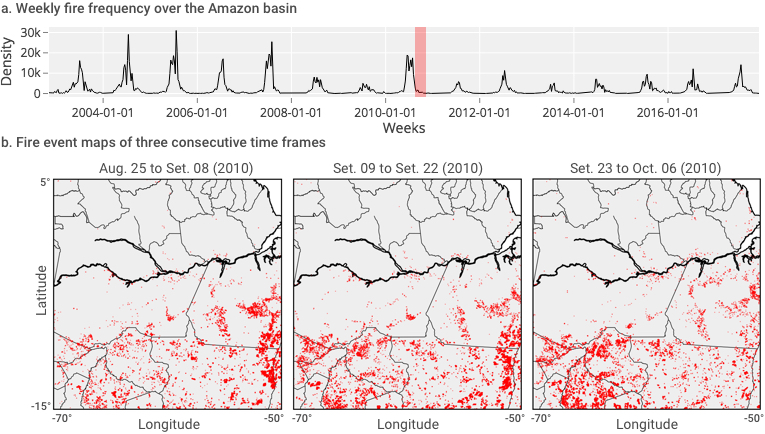}
\caption{\small Fire events over time.}
\label{fig_fireFreq}
\end{figure*}

Our first analysis consists on studying two simple network properties: node degree and node strength. The degree $k_i$ of a node $v_i$ here represents the number of different areas whose fire events occur consecutively with the grid cell of $v_i$. The strength $s_i$ measures the frequency that a fire event occurs consecutively between all the neighbors of $v_i$. Since the degree does not account for frequency, the strength measure seems to bring more information about the networks generated by our method. In Fig \ref{fig_leo3} we present the cumulative strength and degree (inset) distributions for the historical network. Both distributions have a decay faster than a power-law, what suggests the network is not scale-free. These distributions show that many nodes have low degree and low strength. Conversely, the network have just a few nodes with high degree and/or strength. It is important to note that a high degree does not necessarily imply a high strength. But, in fact, the weights and degrees of the historical network are highly correlated ($r = 0.92$, p-value $< 0.01$).

\begin{figure*}[]
\includegraphics[width=0.96\textwidth,height=0.085\textwidth]{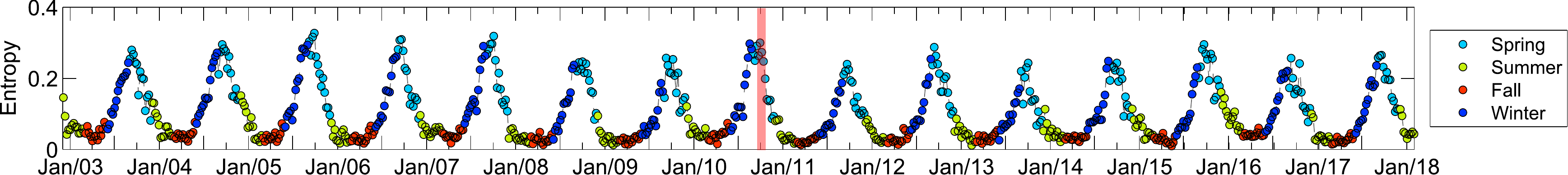}
\caption{\small Normalized entropy of the weekly temporal networks}
\label{fig_entropyTemporal}
\end{figure*}

Since the historical network have just a few nodes with high strength, one next step is to analyze the geographical locations of these nodes. In Figure~\ref{fig_leo4} we illustrate the node strengths for the historical network. The regions with high frequency of consecutive events are similar to those ones found in Figure~\ref{fig_leo1}. However, here it is possible find what are the most frequent sub-regions. In fact, these sub-regions are related to the land use (background colors in Figure~\ref{fig_leo4}). As pointed by a recent research \cite{curtis2018}, the land activity in these regions in the last years are mainly characterized by commodity driven deforestation (agriculture, mining, etc) and shifting agriculture (that is later abandoned). 

\subsection{Temporal network approach}
\label{subsec_temporal}

In the previous section, we explore an approach of constructing a single network from the entire historical period. We observe general patterns related to some macro-regions of fire activity, recognized by a community detection algorithm. In this section, we face the problem form the point of view of multiple layers or temporal networks approach. We generate networks per weeks from the fire events data by employing our proposed construction method. In this way, a set of layers, or temporal networks, represents the spatial fire activities and patterns during each week. The set of networks $\mathcal{G}$ is constructed in consecutive intervals of ${\Delta}t = 7$ days, beginning from Jan. 01, 2003 to Jan. 24, 2018. Formally, $\mathcal{G} = \{G_0, G_1, \ldots, G_{l}\}$ with $l = 786$ layers, where $G_0$ is the network of the first ${\Delta}t$ days, $G_1$ the next ${\Delta}t$ days, and so on.

We start the temporal analysis considering intervals of $7$ days and the frequency of fire events in the grid cells. At first glance in Figure~\ref{fig_fireFreq}(a), we can assume there is a pattern of fire season in the Amazon basin. However, this pattern is not entirely clear over the years and the start and end seasons are not so well defined. As an example, the marked interval from Aug. 25, to Oct. 06, 2010 in Figure~\ref{fig_fireFreq}(a) seems to be at the end of the fire season of 2010. However, in Figure~\ref{fig_fireFreq}(b) we observe there was a high activity of fire events in the basin during this interval of time. Motivated by the lack of precision of the direct frequency approach, we analyze the spatio-temporal fire event data considering weekly temporal networks. For this purpose, we employ the proposed event-occurrence construction method to mine the activity patterns in global (networks), local (nodes) and intermediate (communities) scale.

\begin{figure*}[tb]
\subfigure[Fire frequency]{\includegraphics[width=0.33\textwidth]{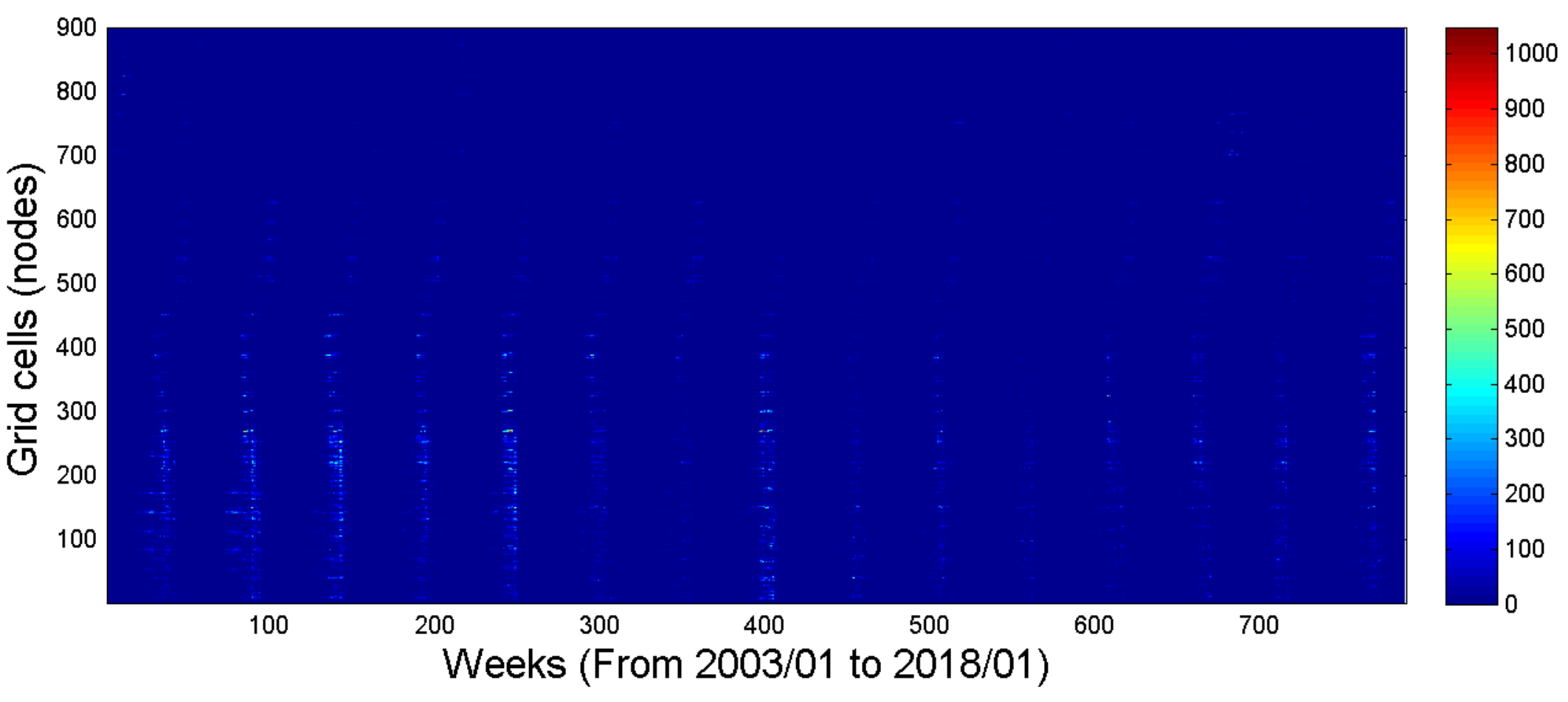}}
\subfigure[Degree centrality]{\includegraphics[width=0.33\textwidth]{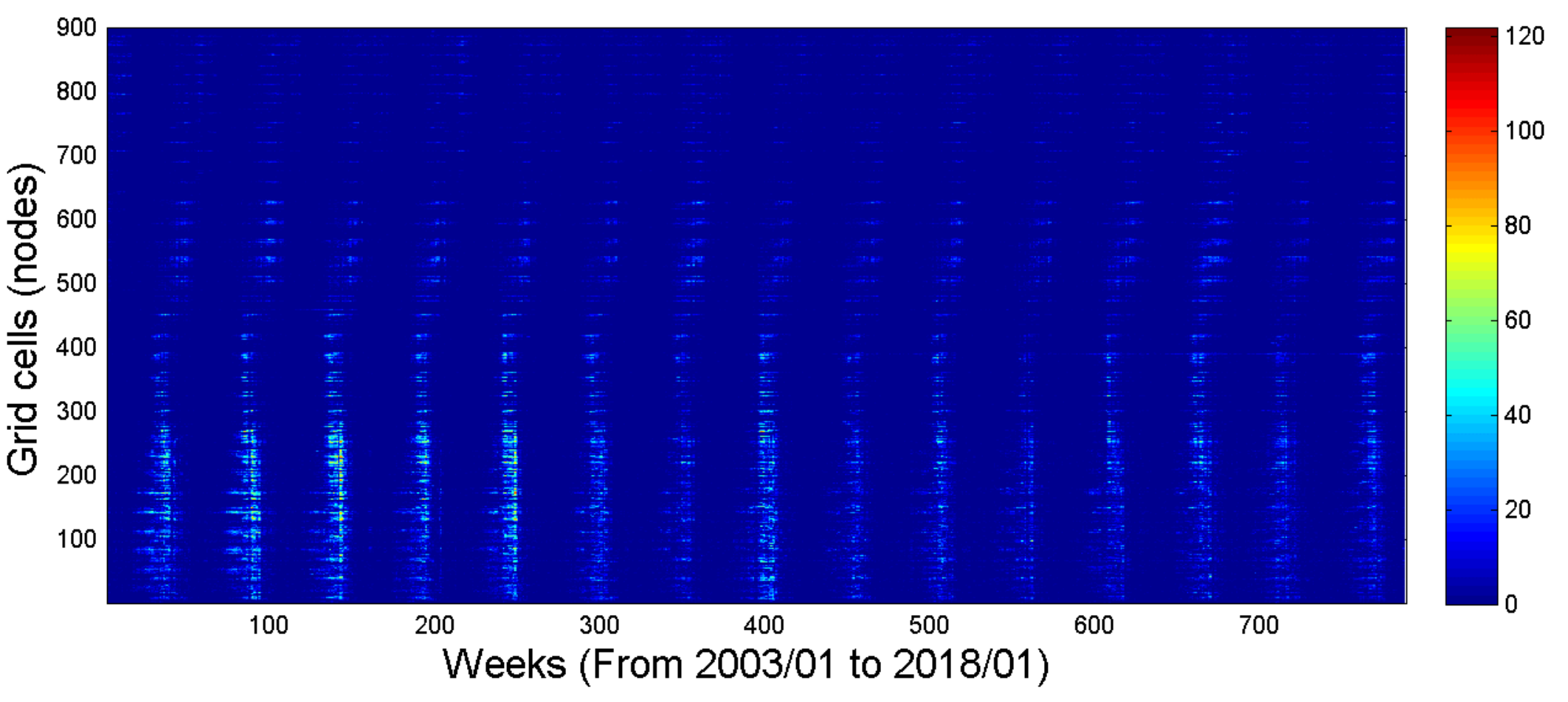}}
\subfigure[K-Core centrality]{\includegraphics[width=0.33\textwidth]{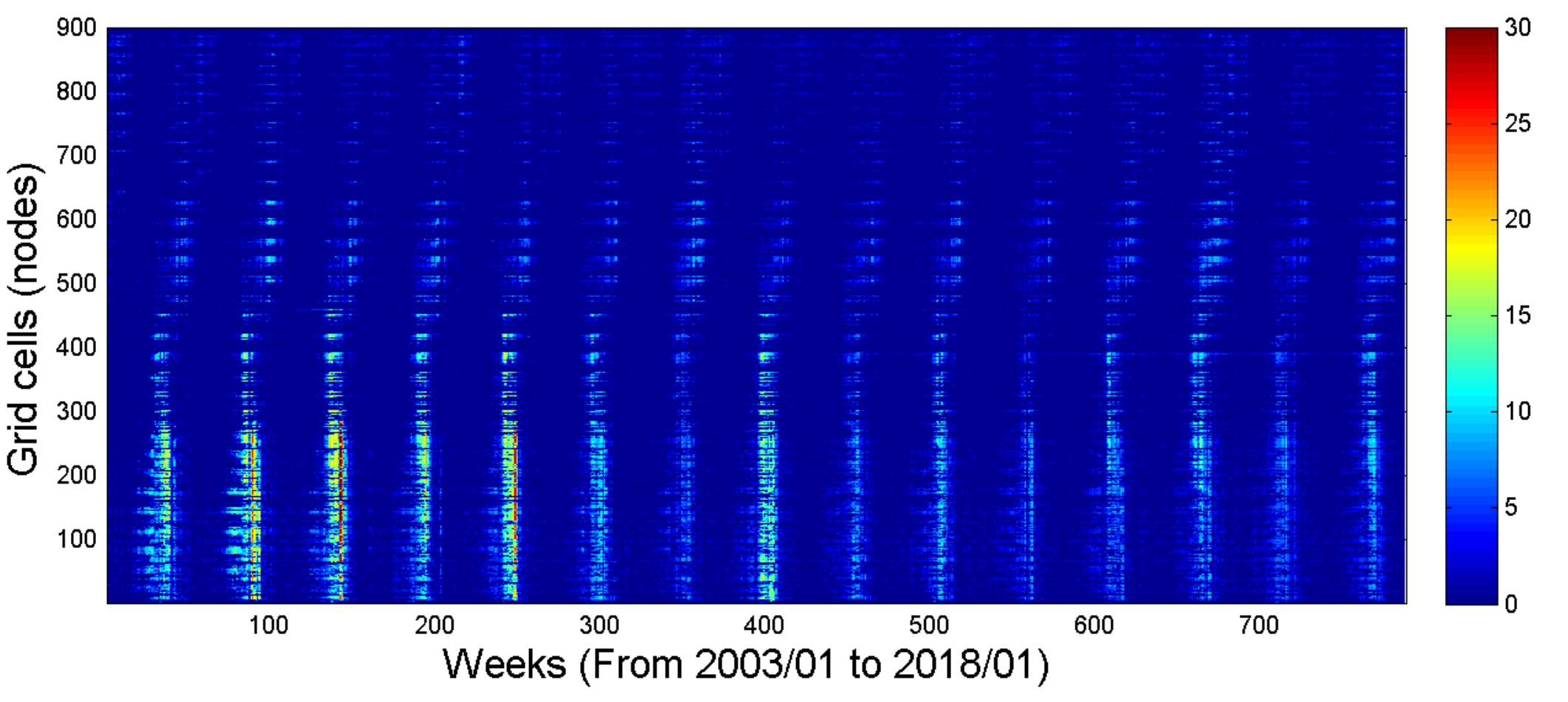}}
\caption{\small (Color online) Comparison of the temporal series values of each node (grid cells) of the studied area per weeks: (a) the fire frequency, (b) the node degree and (c) the K-Core value in the respective weekly temporal network. Higher the frequency or centrality, redder the cell.}
\label{fig_centrality}
\end{figure*}

In a global scale analysis, we explore some general measures to characterize the temporal networks. We calculate the normalized entropy (Eq.~\ref{eq:shannon}) for each network $G_x$ from $\mathcal{G}$, as shown in Figure~\ref{fig_entropyTemporal}, in which higher the entropy, higher the fire events activity and more heterogeneous the network. We can observe a clear pattern of fire activity in the studied region, starting at the beginning of winter and finishing in the middle of summer, which is, in general, the predominant dry season. Different to Figure~\ref{fig_fireFreq}(a) in the marked interval,  the fire activity is in the middle of the peak in Figure~\ref{fig_entropyTemporal}. This result shows the advantages of considering the proposed data modeling method and network mining for obtaining extra knowledge.

Concerning the local scale, we process the information contained in the grid-cells or nodes in the case of networks. The traditional approach is to analyze the activation frequency of the grid cells over time. Then, statistical measures like mean, moments, percentiles,  are used to understand the cells distribution. On the other hand, we can calculate some centrality measures over the network for finding the role of each node in the dynamic evolution. Several topological measures describe the relevance of nodes according to structural and dynamical properties~\cite{Lu2016}. Many have been employed in climatological problems, like in forecasting and prediction~\cite{Fan2017, Meng2018}, disaster risk, and management~\cite{Santos2019}, among others. In particular, the degree and K-Core have great prominence in the area. For instance, K-Core centrality presents a good agreement for identifying the essential nodes in different dynamics, like the most influential spreaders in diffusion process~\cite{Kitsak2010,socinf2015}, the best target nodes for vaccination or marketing campaigns~\cite{Hebert-Dufresne2013, JStatRumor2017}, and other domains.

\subsubsection{Centrality Measures over time}

Figure~\ref{fig_centrality} shows the temporal values per week of each node or grid cell of the Amazon basin. As defined in Section~\ref{subsec_method}, we have a total of $900$ grid cells, where node $0$ is the leftmost cell at the bottom of the grid. We can observe a faint seasonal pattern in the frequency of fire events per weeks for each cell over time (Figure~\ref{fig_centrality}(a)). In particular, for id cells over $500$ (y-axis), it is difficult to see the frequency values, leading to wrongly understand that there is a non-significant fire dynamic in the central and north regions. Besides, for x-axis from week $400$ to $900$, it seems the seasonal pattern tend to vanish. Opposite, when considering the temporal networks, the degree and K-Core centralities (Figure~\ref{fig_centrality}(b) and (c) respectively, with the darker color representing the highest centrality) capture well the seasonal fire pattern of each year. These centrality measures enable to characterize better the role of each node in the fire dynamic, e.g., nodes of the north region (above id $500$ y-axis) have a more clear fire activity in Figures~\ref{fig_centrality}(b) and (c) than Figure~\ref{fig_centrality}(a).

The K-Core measure revealed to be particularly interesting in the context of fire activities. Nodes with higher K-Core centrality are most central in the fire activity, i.e., they are the centroids of the fire events. Nodes in the south (from $0$ to $300$), are the central focus of fire activity in the basin. Also, Figure~\ref{fig_centrality}(c) illustrates a lag pattern in the starting and finishing fire activity between the southern regions (bottom in the Figure) and northern regions (top in the Figure). This result indicates a dynamic of movement or displacement of fire activities in the Amazon basin throughout the year. The fire displacement may occur due to favorable environmental conditions and land use in the regions. The fire season starts at the southern areas and advances to the northern areas through the year.

\subsubsection{The Centrality-series Similarity Network} 

We define the centrality series of a node as the consecutive measurements of its centrality $C$ over time. Formally, for $v_i \in \mathcal{G}$ and the centrality value $C\,(G,v_i)$, with $G$ the network where is calculated, the time series $\chi^{_C}_i$ is the set of sequential centrality values among the temporal networks, i.e., $\chi^{_C}_i = \{C\,(G_0,v_i), C\,(G_1,v_i), \ldots , C\,(G_{l},v_i) \}$, and $l$ the number of temporal networks. $\chi^{_C}_i$ describes the importance of the node in the geographical system according to its centrality evolution. Thus, we aim to understand how the nodes are related and are similar over time. The first step is to calculate the similarity among all the time series of centralities. Then, we generate the centrality-series similarity matrix $\hat{\chi}^{_C}$, defined as follow:

\begin{equation}
\resizebox{0.4\textwidth}{!}{$%
\hat{\chi}^{_C} = 
\begin{pmatrix}
r\,(\chi^{_C}_0,\chi^{_C}_0)  & \dots & r\,(\chi^{_C}_0,\chi^{_C}_j) & \ldots & r\,(\chi^{_C}_0,\chi^{_C}_n) \\
\vdots & \ddots & \vdots  & \vdots & \vdots  \\
r\,(\chi^{_C}_j,\chi^{_C}_0)  & \dots & r\,(\chi^{_C}_j,\chi^{_C}_j) & \ldots & r\,(\chi^{_C}_j,\chi^{_C}_n) \\
\vdots & \vdots & \vdots  & \ddots & \vdots  \\
r\,(\chi^{_C}_n,\chi^{_C}_0)  & \dots & r\,(\chi^{_C}_n,\chi^{_C}_j) & \ldots & r\,(\chi^{_C}_n,\chi^{_C}_n)
\end{pmatrix}
$%
}
\end{equation}

Here, we adopt the similarity function $r$ as the Pearson correlation and the centrality $C = KC$ as the K-Core measure. The full matrix $\hat{\chi}^{_{KC}} \in \Re^{n \times n}$ brings the similarity of temporal K-Core series between pairs of nodes. For the sake of simplicity, we will refer to this centrality-series similarity matrix as ``CSS matrix''. Next, we employ the CSS matrix of K-Core for obtaining the centrality-series similarity \emph{(CSS) network}. The constructed network allows finding local and dense structures according to the temporal centrality similarity of the nodes. For this goal, we employ the well known $\mathbf{k}$NN construction method, where a hub in the data space is a hub in the $\mathbf{k}$NN network~\cite{Berton2018}. The $\mathbf{k}$NN method consists in connect each node with its $\mathbf{k}$ most  similar or nearest neighbors. There exists a link between $v_i$ and $v_j$ if $v_i \in K(\hat{\chi}^{_{KC}}_{j}) \, \vee \, v_j \in K(\hat{\chi}^{_{KC}}_{i})$, where $K(\hat{\chi}^{_{KC}}_{i})$ is the set formed by the $\mathbf{k}$ nearest neighbors of $v_i$, and the main diagonal of $\hat{\chi}^{_{KC}} = 0^{1 \times n}$. The larger the $\mathbf{k}$, the denser the network.

\begin{figure}[]
\includegraphics[width=0.4\textwidth]{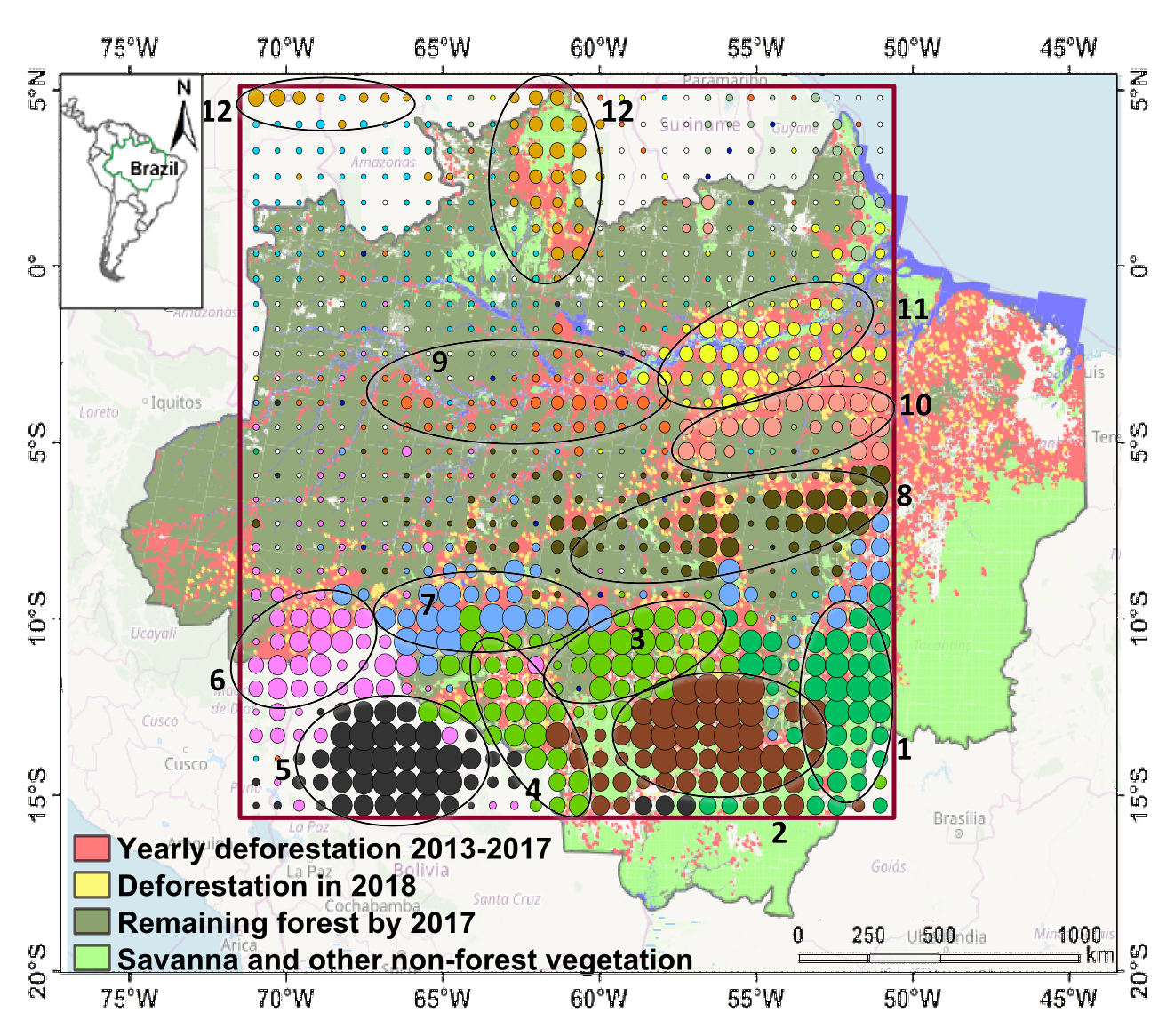}
\caption{\small (Color online) Community structure of the geographic fire events in the Amazon basin. The size of the grid cells represents the overall intensity of fire activity over the time. In the background, deforestation (2013-2017) is shown in red, with yellow point for 2018. Data from 
Instituto Nacional de Pesquisas Espaciais ~\cite{prodesINPE2018}.}
\label{fig_temporalcommunities}
\end{figure}

\begin{figure*}[]
\includegraphics[width=0.85\textwidth,height=0.3\textwidth]{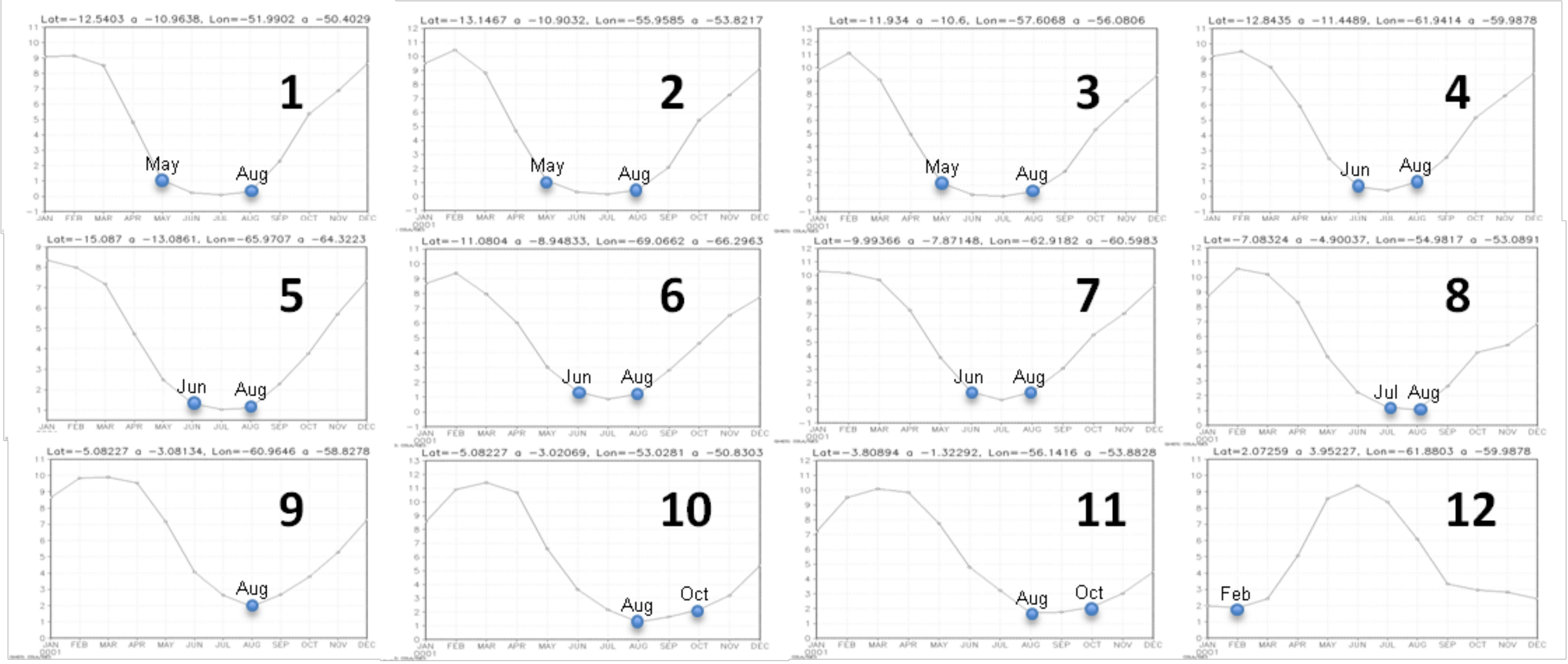}
\caption{\small Average monthly precipitation for the detected communities. Big symbols show the interval of intense dry period.}
\label{fig_dryseason}
\end{figure*}

\subsubsection{Clustering Patterns and Heterogeneous Regions of Fire Events} 

We generate the CSS network using the $\mathbf{k}$NN method with parameter $\mathbf{k} = 3$. Then, we calculate the community structure of the CSS network by the modular community detection method~\cite{Lambiotte2014} (Figure~\ref{fig_temporalcommunities}). We obtain $12$ significant communities, disregarding all the grid cells where the MODIS never reported a fire event, i.e., K-Core equal to zero in all the period. The subregions were analyzed concerning the spatio-temporal patterns and the fire activity. In this way, our method satisfactorily recognizes subregions that present dynamical differences in dry periods and intensity of fire activity. 

We compared the subregions concerning the average monthly precipitation of the CMAP data products from NOAA/ OAR/ ESRL PSD~\footnote{avaliable at \url{https://www.esrl.noaa.gov/psd/}}. This is a global precipitation data set of 30-years monthly analysis based on gauge observations, satellite estimates, and numerical models~\cite{HuffmanRain97}. The precipitation curves of each subregion are depicted in Figure~\ref{fig_dryseason}. The dry season starts with values lower than $2.0$ mm/day, and are marked in the curves. We find the following patterns concerning the precipitation: 
{\bf (\textit{i})} Subregions 4-7 have dry seasons occurring approximately in the same period: 4 and 5 have less precipitation than 6 and 7, especially in the dry season.
{\bf (\textit{ii})} Subregions 2 and 3 have the longest and more severe dry season than 4-7; 1 has a similar pattern to 2-3, but less precipitation in total (values in the rainy season are lower).
{\bf (\textit{iii})} Subregions 8-11 have a rainy season starting later and better defined (fewer months with high precipitation). Consequently, the dry season ends later: The rainy season of 9 also starts later but its dry season is a little shorter than 8,10,11.
{\bf (\textit{iv})} Subregion 12 presents the least intense rainy season (lower maximum values) and starts later. Consequently, the dry season is also relatively long and least intense. Subregion 12 has different period of the dry season than all other regions.

Finally, about vegetation cover and land use, the subregions have the following characterization: {\bf (\textit{i})} Subregions 1-4 are related to land use related to pastures in old areas of deforestation, with  subregion 3 relatively old. Subregion 2 has land use more related to agriculture, pasture, and areas that were forests. {\bf (\textit{ii})} Subregion 5 is related to a savanna region in Bolivia. Due to the high fire activity detected in 5, there must be land use with agriculture. {\bf (\textit{iii})} Subregions 6-8,10 are related to deforestation in areas surrounding reserves and roadsides with more strong gradients of forests and recently deforested areas. Subregion 8 should be related to more recent deforestation, with pasture and agriculture. Subregion 6 is related to land use with deforestation near roads. {\bf (\textit{iv})} Subregions 9 and 11 are areas near the Amazon River. They are more related to recent settlements where there must be deforestation, agriculture, and pasture. {\bf (\textit{v})} Subregion 12 is a savanna region surrounded by forest with deforestation and land use. It is a combination of natural propensity to fire with pasture maintenance.

\section{Final remarks and Future works}\label{sec:finalRemarks}

In this paper, we explored the fire activity in the Amazon basin as a complex system. We presented a data modeling method for constructing chronological networks (Sec.~\ref{subsec_method}) and two graph mining approaches (Sec.~\ref{subsec_single_net} and \ref{subsec_temporal}). We observed that these approaches captured important insights hidden in the fire data that other network-based models are incapable of highlighting. As results, we found subregions -- or communities -- with particular differences in the temporal (by its dry season periods), activity (by the intensity of fire events) and space (by the land-use and location)  conditions for the occurrence of fire. Moreover, a pattern of fire-activity displacement over the year, starting from the southeastern subregions, spreading in the south and rising to the northwest of the basin was also identified. Also, although both approaches follow different ways to analyze the network, they recognize the same set of most active cells (Fig.~\ref{fig_leo3} and Fig.~\ref{fig_temporalcommunities}). In summary, our method allows the study of spatio-temporal data sets from a different perspective.

For future works, we intend to employ and propose advanced techniques that could lead to more revelations on this kind of spatio-temporal problems. Evaluating more sophisticated rules, like linking nodes by considering a spatial threshold, is a natural path of this work for understanding how the new rules could affect the network structure. Another important point is that the same fixed time window does not necessarily space the events. Adding a memory-mechanism that avoid the time slicing approach, but allows to measure changes over time, could be a new strategy to understand temporal changes in complex systems.

\begin{acks}
This research is supported by the \grantsponsor{SP3255}{Funda\c{c}\~{a}o de Amparo \`{a} Pesquisa do Estado de S\~{a}o Paulo (FAPESP)}{https://bv.fapesp.br/en/auxilios/93229/dynamic-phenomena-in-complex-networks-basics-and-applications/} under Grant
No.: \grantnum{SP3255}{2015/50122-0} and the \grantsponsor{SP3255}{German Research Council (DFG-GRTK)}{http://www.dfg.de/en/funded_projects/current_projects_programmes/list/projectdetails/index.jsp?id=183049896} Grant No.: \grantnum{SP250}{1740/2}. D.A.V.O acknowledges FAPESP (Grants \grantnum{SP3255}{2016/23698-1}, \grantnum{SP3255}{2018/01722-3}, and \grantnum{SP3255}{2018/24260-5}). L. N. F. thanks FAPESP under Grant No.:~\grantnum{SP3255}{2017/05831-9} and \grantsponsor{SP4414}{Brazilian Higher Education Funding Council (CAPES)}{http://www.capes.gov.br/} scholarship.

\end{acks}

%% file: sigconf.bbl

\begin{thebibliography}{30}


\ifx \showCODEN    \undefined \def \showCODEN     #1{\unskip}     \fi
\ifx \showDOI      \undefined \def \showDOI       #1{#1}\fi
\ifx \showISBNx    \undefined \def \showISBNx     #1{\unskip}     \fi
\ifx \showISBNxiii \undefined \def \showISBNxiii  #1{\unskip}     \fi
\ifx \showISSN     \undefined \def \showISSN      #1{\unskip}     \fi
\ifx \showLCCN     \undefined \def \showLCCN      #1{\unskip}     \fi
\ifx \shownote     \undefined \def \shownote      #1{#1}          \fi
\ifx \showarticletitle \undefined \def \showarticletitle #1{#1}   \fi
\ifx \showURL      \undefined \def \showURL       {\relax}        \fi
\providecommand\bibfield[2]{#2}
\providecommand\bibinfo[2]{#2}
\providecommand\natexlab[1]{#1}
\providecommand\showeprint[2][]{arXiv:#2}

\bibitem[\protect\citeauthoryear{Abe and Suzuki}{Abe and Suzuki}{2006}]%
        {Abe2006}
\bibfield{author}{\bibinfo{person}{S. Abe} {and} \bibinfo{person}{N. Suzuki}.}
  \bibinfo{year}{2006}\natexlab{}.
\newblock \showarticletitle{{Complex-network description of seismicity}}.
\newblock \bibinfo{journal}{\emph{Nonlinear Processes in Geophysics}}
  \bibinfo{volume}{13}, \bibinfo{number}{2} (\bibinfo{date}{may}
  \bibinfo{year}{2006}), \bibinfo{pages}{145--150}.
\newblock


\bibitem[\protect\citeauthoryear{Berton, de~Andrade~Lopes, and
  Vega-Oliveros}{Berton et~al\mbox{.}}{2018}]%
        {Berton2018}
\bibfield{author}{\bibinfo{person}{L. Berton}, \bibinfo{person}{A. de
  Andrade~Lopes}, {and} \bibinfo{person}{D.~A. Vega-Oliveros}.}
  \bibinfo{year}{2018}\natexlab{}.
\newblock \showarticletitle{A Comparison of Graph Construction Methods for
  Semi-Supervised Learning}. In \bibinfo{booktitle}{\emph{2018 International
  Joint Conference on Neural Networks (IJCNN)}}. \bibinfo{pages}{1--8}.
\newblock
\showISSN{2161-4407}


\bibitem[\protect\citeauthoryear{Bialonski, Wendler, and Lehnertz}{Bialonski
  et~al\mbox{.}}{2011}]%
        {Bialonski2011}
\bibfield{author}{\bibinfo{person}{Stephan Bialonski}, \bibinfo{person}{Martin
  Wendler}, {and} \bibinfo{person}{Klaus Lehnertz}.}
  \bibinfo{year}{2011}\natexlab{}.
\newblock \showarticletitle{{Unraveling Spurious Properties of Interaction
  Networks with Tailored Random Networks}}.
\newblock \bibinfo{journal}{\emph{PLoS ONE}} \bibinfo{volume}{6},
  \bibinfo{number}{8} (\bibinfo{date}{aug} \bibinfo{year}{2011}),
  \bibinfo{pages}{e22826}.
\newblock
\showISSN{1932-6203}


\bibitem[\protect\citeauthoryear{Brando, Balch, Nepstad, Morton, Putz, Coe,
  Silv{\'{e}}rio, Macedo, Davidson, N{\'{o}}brega, Alencar, and
  Soares-Filho}{Brando et~al\mbox{.}}{2014}]%
        {Brando2014}
\bibfield{author}{\bibinfo{person}{P.~M. Brando}, \bibinfo{person}{J.~K Balch},
  \bibinfo{person}{D.~C Nepstad}, \bibinfo{person}{D.~C Morton},
  \bibinfo{person}{F.~E. Putz}, \bibinfo{person}{M.~T Coe}, \bibinfo{person}{D.
  Silv{\'{e}}rio}, \bibinfo{person}{M.~N. Macedo}, \bibinfo{person}{E.~A.
  Davidson}, \bibinfo{person}{C.~C. N{\'{o}}brega}, \bibinfo{person}{A.
  Alencar}, {and} \bibinfo{person}{B.~S. Soares-Filho}.}
  \bibinfo{year}{2014}\natexlab{}.
\newblock \showarticletitle{{Abrupt increases in Amazonian tree mortality due
  to drought-fire interactions.}}
\newblock \bibinfo{journal}{\emph{Proceedings of the National Academy of
  Sciences of the United States of America}} \bibinfo{volume}{111},
  \bibinfo{number}{17} (\bibinfo{date}{apr} \bibinfo{year}{2014}),
  \bibinfo{pages}{6347--52}.
\newblock
\showISSN{1091-6490}


\bibitem[\protect\citeauthoryear{Curtis, Slay, Harris, Tyukavina, and
  Hansen}{Curtis et~al\mbox{.}}{2018}]%
        {curtis2018}
\bibfield{author}{\bibinfo{person}{Philip~G. Curtis},
  \bibinfo{person}{Christy~M. Slay}, \bibinfo{person}{Nancy~L. Harris},
  \bibinfo{person}{Alexandra Tyukavina}, {and} \bibinfo{person}{Matthew~C.
  Hansen}.} \bibinfo{year}{2018}\natexlab{}.
\newblock \showarticletitle{Classifying drivers of global forest loss}.
\newblock \bibinfo{journal}{\emph{Science}} \bibinfo{volume}{361},
  \bibinfo{number}{6407} (\bibinfo{year}{2018}), \bibinfo{pages}{1108--1111}.
\newblock
\showISSN{0036-8075}


\bibitem[\protect\citeauthoryear{Dey and Schweitzer}{Dey and
  Schweitzer}{2018}]%
        {Dey2018}
\bibfield{author}{\bibinfo{person}{Daniel Dey} {and} \bibinfo{person}{Callie
  Schweitzer}.} \bibinfo{year}{2018}\natexlab{}.
\newblock \showarticletitle{{A Review on the Dynamics of Prescribed Fire, Tree
  Mortality, and Injury in Managing Oak Natural Communities to Minimize
  Economic Loss in North America}}.
\newblock \bibinfo{journal}{\emph{Forests}} \bibinfo{volume}{9},
  \bibinfo{number}{8} (\bibinfo{date}{jul} \bibinfo{year}{2018}),
  \bibinfo{pages}{461}.
\newblock
\showISSN{1999-4907}


\bibitem[\protect\citeauthoryear{Fan, Meng, Ashkenazy, Havlin, and
  Schellnhuber}{Fan et~al\mbox{.}}{2017}]%
        {Fan2017}
\bibfield{author}{\bibinfo{person}{Jingfang Fan}, \bibinfo{person}{Jun Meng},
  \bibinfo{person}{Yosef Ashkenazy}, \bibinfo{person}{Shlomo Havlin}, {and}
  \bibinfo{person}{Hans~Joachim Schellnhuber}.}
  \bibinfo{year}{2017}\natexlab{}.
\newblock \showarticletitle{{Network analysis reveals strongly localized
  impacts of El Ni{\~{n}}o.}}
\newblock \bibinfo{journal}{\emph{Proceedings of the National Academy of
  Sciences of the United States of America}} \bibinfo{volume}{114},
  \bibinfo{number}{29} (\bibinfo{date}{jul} \bibinfo{year}{2017}),
  \bibinfo{pages}{7543--7548}.
\newblock
\showISSN{1091-6490}


\bibitem[\protect\citeauthoryear{Farkas, Jeong, Vicsek, Barab{\'{a}}si, and
  Oltvai}{Farkas et~al\mbox{.}}{2003}]%
        {Farkas2003}
\bibfield{author}{\bibinfo{person}{I. Farkas}, \bibinfo{person}{H. Jeong},
  \bibinfo{person}{T. Vicsek}, \bibinfo{person}{A.-L. Barab{\'{a}}si}, {and}
  \bibinfo{person}{Z.N. Oltvai}.} \bibinfo{year}{2003}\natexlab{}.
\newblock \showarticletitle{{The topology of the transcription regulatory
  network in the yeast, Saccharomyces cerevisiae}}.
\newblock \bibinfo{journal}{\emph{Physica A: Statistical Mechanics and its
  Applications}} \bibinfo{volume}{318}, \bibinfo{number}{3-4}
  (\bibinfo{date}{feb} \bibinfo{year}{2003}), \bibinfo{pages}{601--612}.
\newblock
\showISSN{0378-4371}


\bibitem[\protect\citeauthoryear{Ferreira, Ribeiro, Papa, and Menezes}{Ferreira
  et~al\mbox{.}}{2018}]%
        {Ferreira2018}
\bibfield{author}{\bibinfo{person}{Douglas Ferreira}, \bibinfo{person}{Jennifer
  Ribeiro}, \bibinfo{person}{Andr{\'{e}}s Papa}, {and} \bibinfo{person}{Ronaldo
  Menezes}.} \bibinfo{year}{2018}\natexlab{}.
\newblock \showarticletitle{{Towards evidence of long-range correlations in
  shallow seismic activities}}.
\newblock \bibinfo{journal}{\emph{EPL}} \bibinfo{volume}{121},
  \bibinfo{number}{5} (\bibinfo{date}{mar} \bibinfo{year}{2018}),
  \bibinfo{pages}{58003}.
\newblock
\showISSN{12864854}


\bibitem[\protect\citeauthoryear{Fortunato and Hric}{Fortunato and
  Hric}{2016}]%
        {Fortunato16}
\bibfield{author}{\bibinfo{person}{Santo Fortunato} {and}
  \bibinfo{person}{Darko Hric}.} \bibinfo{year}{2016}\natexlab{}.
\newblock \showarticletitle{Community detection in networks: A user guide}.
\newblock \bibinfo{journal}{\emph{Physics Reports}}  \bibinfo{volume}{659}
  (\bibinfo{year}{2016}), \bibinfo{pages}{1--44}.
\newblock


\bibitem[\protect\citeauthoryear{Gatti, Gloor, Miller, Doughty, Malhi,
  Domingues, Basso, Martinewski, Correia, Borges, Freitas, Braz, Anderson,
  Rocha, Grace, Phillips, and Lloyd}{Gatti et~al\mbox{.}}{2014}]%
        {Gatti2014}
\bibfield{author}{\bibinfo{person}{L.~V. Gatti}, \bibinfo{person}{M. Gloor},
  \bibinfo{person}{J.~B. Miller}, \bibinfo{person}{C.~E. Doughty},
  \bibinfo{person}{Y. Malhi}, \bibinfo{person}{L.~G. Domingues},
  \bibinfo{person}{L.~S. Basso}, \bibinfo{person}{A. Martinewski},
  \bibinfo{person}{C.~S.~C. Correia}, \bibinfo{person}{V.~F. Borges},
  \bibinfo{person}{S. Freitas}, \bibinfo{person}{R. Braz},
  \bibinfo{person}{L.~O. Anderson}, \bibinfo{person}{H. Rocha},
  \bibinfo{person}{J. Grace}, \bibinfo{person}{O.~L. Phillips}, {and}
  \bibinfo{person}{J. Lloyd}.} \bibinfo{year}{2014}\natexlab{}.
\newblock \showarticletitle{{Drought sensitivity of Amazonian carbon balance
  revealed by atmospheric measurements}}.
\newblock \bibinfo{journal}{\emph{Nature}} \bibinfo{volume}{506},
  \bibinfo{number}{7486} (\bibinfo{date}{feb} \bibinfo{year}{2014}),
  \bibinfo{pages}{76--80}.
\newblock
\showISSN{0028-0836}


\bibitem[\protect\citeauthoryear{H{\'{e}}bert-Dufresne, Allard, Young, and
  Dub{\'{e}}}{H{\'{e}}bert-Dufresne et~al\mbox{.}}{2013}]%
        {Hebert-Dufresne2013}
\bibfield{author}{\bibinfo{person}{Laurent H{\'{e}}bert-Dufresne},
  \bibinfo{person}{Antoine Allard}, \bibinfo{person}{Jean-Gabriel Young}, {and}
  \bibinfo{person}{Louis~J Dub{\'{e}}}.} \bibinfo{year}{2013}\natexlab{}.
\newblock \showarticletitle{{Global efficiency of local immunization on complex
  networks.}}
\newblock \bibinfo{journal}{\emph{Scientific reports}}  \bibinfo{volume}{3}
  (\bibinfo{date}{jan} \bibinfo{year}{2013}), \bibinfo{pages}{2171}.
\newblock
\showISSN{2045-2322}


\bibitem[\protect\citeauthoryear{Huffman, Adler, Arkin, Chang, Ferraro, Gruber,
  Janowiak, McNab, Rudolf, and Schneider}{Huffman et~al\mbox{.}}{1997}]%
        {HuffmanRain97}
\bibfield{author}{\bibinfo{person}{George~J. Huffman},
  \bibinfo{person}{Robert~F. Adler}, \bibinfo{person}{Philip Arkin},
  \bibinfo{person}{Alfred Chang}, \bibinfo{person}{Ralph Ferraro},
  \bibinfo{person}{Arnold Gruber}, \bibinfo{person}{John Janowiak},
  \bibinfo{person}{Alan McNab}, \bibinfo{person}{Bruno Rudolf}, {and}
  \bibinfo{person}{Udo Schneider}.} \bibinfo{year}{1997}\natexlab{}.
\newblock \showarticletitle{The Global Precipitation Climatology Project (GPCP)
  Combined Precipitation Dataset}.
\newblock \bibinfo{journal}{\emph{Bulletin of the American Meteorological
  Society}} \bibinfo{volume}{78}, \bibinfo{number}{1} (\bibinfo{year}{1997}),
  \bibinfo{pages}{5--20}.
\newblock


\bibitem[\protect\citeauthoryear{INPE}{INPE}{2018}]%
        {prodesINPE2018}
\bibfield{author}{\bibinfo{person}{INPE}.} \bibinfo{year}{2018}\natexlab{}.
\newblock \showarticletitle{Projeto PRODES: Monitoramento da Floresta
  Amazônica Brasileira por Satélite}.
\newblock \bibinfo{journal}{\emph{Instituto Nacional de Pesquisas Espaciais,}}
  \bibinfo{volume}{S\~ao Jos\'e dos Campos, Brazil}, \bibinfo{number}{retrieved
  from \url{http://www.obt.inpe.br/prodes/}} (\bibinfo{year}{2018}).
\newblock


\bibitem[\protect\citeauthoryear{Kitsak, Gallos, Havlin, Liljeros, Muchnik,
  Stanley, and Makse}{Kitsak et~al\mbox{.}}{2010}]%
        {Kitsak2010}
\bibfield{author}{\bibinfo{person}{M. Kitsak}, \bibinfo{person}{L.K. Gallos},
  \bibinfo{person}{S. Havlin}, \bibinfo{person}{F. Liljeros},
  \bibinfo{person}{L. Muchnik}, \bibinfo{person}{H.E. Stanley}, {and}
  \bibinfo{person}{A. Makse}.} \bibinfo{year}{2010}\natexlab{}.
\newblock \showarticletitle{{I}dentification of Influential Spreaders in
  Complex Networks}.
\newblock \bibinfo{journal}{\emph{Nature Physics}} \bibinfo{volume}{6},
  \bibinfo{number}{11} (\bibinfo{year}{2010}), \bibinfo{pages}{888--893}.
\newblock


\bibitem[\protect\citeauthoryear{Lambiotte, Delvenne, and Barahona}{Lambiotte
  et~al\mbox{.}}{2014}]%
        {Lambiotte2014}
\bibfield{author}{\bibinfo{person}{R. Lambiotte}, \bibinfo{person}{J.
  Delvenne}, {and} \bibinfo{person}{M. Barahona}.}
  \bibinfo{year}{2014}\natexlab{}.
\newblock \showarticletitle{Random Walks, Markov Processes and the Multiscale
  Modular Organization of Complex Networks}.
\newblock \bibinfo{journal}{\emph{IEEE Transactions on Network Science and
  Engineering}} \bibinfo{volume}{1}, \bibinfo{number}{2} (\bibinfo{date}{July}
  \bibinfo{year}{2014}), \bibinfo{pages}{76--90}.
\newblock
\showISSN{2327-4697}


\bibitem[\protect\citeauthoryear{L{\"{u}}, Chen, Ren, Zhang, Zhang, and
  Zhou}{L{\"{u}} et~al\mbox{.}}{2016}]%
        {Lu2016}
\bibfield{author}{\bibinfo{person}{Linyuan L{\"{u}}}, \bibinfo{person}{Duanbing
  Chen}, \bibinfo{person}{Xiao-Long Ren}, \bibinfo{person}{Qian-Ming Zhang},
  \bibinfo{person}{Yi-Cheng Zhang}, {and} \bibinfo{person}{Tao Zhou}.}
  \bibinfo{year}{2016}\natexlab{}.
\newblock \showarticletitle{{Vital nodes identification in complex networks}}.
\newblock \bibinfo{journal}{\emph{Physics Reports}}  \bibinfo{volume}{650}
  (\bibinfo{year}{2016}), \bibinfo{pages}{1--63}.
\newblock


\bibitem[\protect\citeauthoryear{Meng, Fan, Ashkenazy, Bunde, and Havlin}{Meng
  et~al\mbox{.}}{2018}]%
        {Meng2018}
\bibfield{author}{\bibinfo{person}{Jun Meng}, \bibinfo{person}{Jingfang Fan},
  \bibinfo{person}{Yosef Ashkenazy}, \bibinfo{person}{Armin Bunde}, {and}
  \bibinfo{person}{Shlomo Havlin}.} \bibinfo{year}{2018}\natexlab{}.
\newblock \showarticletitle{{Forecasting the magnitude and onset of El
  Ni{\~{n}}o based on climate network}}.
\newblock \bibinfo{journal}{\emph{New Journal of Physics}}
  \bibinfo{volume}{20}, \bibinfo{number}{4} (\bibinfo{date}{apr}
  \bibinfo{year}{2018}), \bibinfo{pages}{043036}.
\newblock
\showISSN{1367-2630}


\bibitem[\protect\citeauthoryear{Newman}{Newman}{2010}]%
        {Newman2010}
\bibfield{author}{\bibinfo{person}{Mark Newman}.}
  \bibinfo{year}{2010}\natexlab{}.
\newblock \bibinfo{booktitle}{\emph{Networks: An Introduction}}.
\newblock \bibinfo{publisher}{Oxford University Press, Inc.},
  \bibinfo{address}{New York, NY, USA}.
\newblock


\bibitem[\protect\citeauthoryear{Newman}{Newman}{2004}]%
        {newman2004}
\bibfield{author}{\bibinfo{person}{M~E~J Newman}.}
  \bibinfo{year}{2004}\natexlab{}.
\newblock \showarticletitle{{F}ast algorithm for detecting community structure
  in networks}.
\newblock \bibinfo{journal}{\emph{Physical Review E}} \bibinfo{volume}{69},
  \bibinfo{number}{3} (\bibinfo{year}{2004}), \bibinfo{pages}{66133}.
\newblock


\bibitem[\protect\citeauthoryear{Rogers}{Rogers}{2018}]%
        {AdamRogers}
\bibfield{author}{\bibinfo{person}{Adam Rogers}.}
  \bibinfo{year}{2018}\natexlab{}.
\newblock \showarticletitle{The Only Thing Fire Scientists Are Sure of: This
  Will Get Worse}.
\newblock \bibinfo{journal}{\emph{WIRED}} (\bibinfo{date}{Aug, 1,}
  \bibinfo{year}{2018}), \bibinfo{pages}{retrieved from
  \url{https://www.wired.com}}.
\newblock


\bibitem[\protect\citeauthoryear{Santos, Londe, Carvalho, Menasch\'e, and
  Vega-Oliveros}{Santos et~al\mbox{.}}{2019}]%
        {Santos2019}
\bibfield{author}{\bibinfo{person}{Leonardo B.~L. Santos},
  \bibinfo{person}{Luciana~R. Londe}, \bibinfo{person}{Tiago Carvalho},
  \bibinfo{person}{Daniel~S. Menasch\'e}, {and} \bibinfo{person}{Didier~A.
  Vega-Oliveros}.} \bibinfo{year}{2019}\natexlab{}.
\newblock \bibinfo{booktitle}{\emph{About interfaces between Machine Learning,
  Complex Networks, Survivability Analysis and Disaster Risk Reduction}}.
\newblock \bibinfo{publisher}{Springer International Publishing}, Chapter In
  press, \bibinfo{pages}{1–32}.
\newblock


\bibitem[\protect\citeauthoryear{Seidman}{Seidman}{1983}]%
        {kcore:seidman83}
\bibfield{author}{\bibinfo{person}{S.B. Seidman}.}
  \bibinfo{year}{1983}\natexlab{}.
\newblock \showarticletitle{Network structure and minimum degree}.
\newblock \bibinfo{journal}{\emph{Social Networks}} \bibinfo{volume}{5},
  \bibinfo{number}{3} (\bibinfo{year}{1983}).
\newblock


\bibitem[\protect\citeauthoryear{Tsonis and Swanson}{Tsonis and
  Swanson}{2008}]%
        {Tsonis2008}
\bibfield{author}{\bibinfo{person}{Anastasios~A. Tsonis} {and}
  \bibinfo{person}{Kyle~L. Swanson}.} \bibinfo{year}{2008}\natexlab{}.
\newblock \showarticletitle{{Topology and predictability of El Ni{\~{n}}o and
  la Ni{\~{n}}a Networks}}.
\newblock \bibinfo{journal}{\emph{Physical Review Letters}}
  \bibinfo{volume}{100}, \bibinfo{number}{22} (\bibinfo{date}{jun}
  \bibinfo{year}{2008}), \bibinfo{pages}{228502}.
\newblock
\showISBNx{0031-9007}
\showISSN{00319007}


\bibitem[\protect\citeauthoryear{Vega{-}Oliveros, Berton, Lopes, and
  Rodrigues}{Vega{-}Oliveros et~al\mbox{.}}{2015}]%
        {socinf2015}
\bibfield{author}{\bibinfo{person}{D. Vega{-}Oliveros}, \bibinfo{person}{L.
  Berton}, \bibinfo{person}{A. Lopes}, {and} \bibinfo{person}{F. Rodrigues}.}
  \bibinfo{year}{2015}\natexlab{}.
\newblock \showarticletitle{Influence Maximization Based on the Least
  Influential Spreaders}. In \bibinfo{booktitle}{\emph{SocInf 2015, co-located
  with {IJCAI} 2015}}, Vol.~\bibinfo{volume}{1398}. \bibinfo{pages}{3--8}.
\newblock


\bibitem[\protect\citeauthoryear{Vega-Oliveros, da~F~Costa, and
  Rodrigues}{Vega-Oliveros et~al\mbox{.}}{2017}]%
        {JStatRumor2017}
\bibfield{author}{\bibinfo{person}{Didier~A Vega-Oliveros},
  \bibinfo{person}{Luciano da F~Costa}, {and} \bibinfo{person}{Francisco~A
  Rodrigues}.} \bibinfo{year}{2017}\natexlab{}.
\newblock \showarticletitle{Rumor propagation with heterogeneous transmission
  in social networks}.
\newblock \bibinfo{journal}{\emph{Journal of Statistical Mechanics: Theory and
  Experiment}} \bibinfo{volume}{2017}, \bibinfo{number}{2}
  (\bibinfo{year}{2017}), \bibinfo{pages}{023401}.
\newblock


\bibitem[\protect\citeauthoryear{Vieira, Toledo, Silva, and Higuchi}{Vieira
  et~al\mbox{.}}{2008}]%
        {Vieira2008}
\bibfield{author}{\bibinfo{person}{ICG. Vieira}, \bibinfo{person}{PM. Toledo},
  \bibinfo{person}{JMC. Silva}, {and} \bibinfo{person}{H. Higuchi}.}
  \bibinfo{year}{2008}\natexlab{}.
\newblock \showarticletitle{{Deforestation and threats to the biodiversity of
  Amazonia}}.
\newblock \bibinfo{journal}{\emph{Brazilian Journal of Biology}}
  \bibinfo{volume}{68}, \bibinfo{number}{4 suppl} (\bibinfo{date}{nov}
  \bibinfo{year}{2008}), \bibinfo{pages}{949--956}.
\newblock
\showISSN{1519-6984}


\bibitem[\protect\citeauthoryear{Wang, Tang, Guo, and Xiu}{Wang
  et~al\mbox{.}}{2006}]%
        {wang06}
\bibfield{author}{\bibinfo{person}{Bing Wang}, \bibinfo{person}{Huanwen Tang},
  \bibinfo{person}{Chonghui Guo}, {and} \bibinfo{person}{Zhilong Xiu}.}
  \bibinfo{year}{2006}\natexlab{}.
\newblock \showarticletitle{{Entropy optimization of scale-free networks'
  robustness to random failures}}.
\newblock \bibinfo{journal}{\emph{Physica A: Statistical Mechanics and its
  Applications}} \bibinfo{volume}{363}, \bibinfo{number}{2}
  (\bibinfo{year}{2006}), \bibinfo{pages}{591--596}.
\newblock


\bibitem[\protect\citeauthoryear{Zemp, Schleussner, Barbosa, and Rammig}{Zemp
  et~al\mbox{.}}{2017}]%
        {Zemp2017}
\bibfield{author}{\bibinfo{person}{D.~C. Zemp}, \bibinfo{person}{C.-F.
  Schleussner}, \bibinfo{person}{H.~M.~J. Barbosa}, {and} \bibinfo{person}{A.
  Rammig}.} \bibinfo{year}{2017}\natexlab{}.
\newblock \showarticletitle{{Deforestation effects on Amazon forest
  resilience}}.
\newblock \bibinfo{journal}{\emph{Geophysical Research Letters}}
  \bibinfo{volume}{44}, \bibinfo{number}{12} (\bibinfo{date}{jun}
  \bibinfo{year}{2017}), \bibinfo{pages}{6182--6190}.
\newblock
\showISSN{00948276}


\bibitem[\protect\citeauthoryear{Zhou, Gozolchiani, Ashkenazy, and Havlin}{Zhou
  et~al\mbox{.}}{2015}]%
        {Zhou2015}
\bibfield{author}{\bibinfo{person}{Dong Zhou}, \bibinfo{person}{Avi
  Gozolchiani}, \bibinfo{person}{Yosef Ashkenazy}, {and}
  \bibinfo{person}{Shlomo Havlin}.} \bibinfo{year}{2015}\natexlab{}.
\newblock \showarticletitle{{Teleconnection Paths via Climate Network Direct
  Link Detection}}.
\newblock \bibinfo{journal}{\emph{Physical Review Letters}}
  \bibinfo{volume}{115}, \bibinfo{number}{26} (\bibinfo{date}{dec}
  \bibinfo{year}{2015}), \bibinfo{pages}{268501}.
\newblock
\showISSN{0031-9007}


\end{thebibliography}
